\documentclass{article}
\usepackage{tabularx}
\usepackage{array}
\newcolumntype{Y}{>{\raggedright\arraybackslash}X}
\usepackage{PRIMEarxiv}
\usepackage{amsmath} 
\usepackage[utf8]{inputenc} 
\usepackage[T1]{fontenc}    
\usepackage[hidelinks]{hyperref}       
\usepackage{url}            
\usepackage{booktabs}       
\usepackage{amsfonts}       
\usepackage{nicefrac}       
\usepackage{microtype}      
\usepackage{lipsum}
\usepackage{fancyhdr}       
\usepackage{graphicx}       
\graphicspath{{media/}}     

\title{Gravity-Aware Hierarchical Routing for Lightweight SensorLLM on Human Activity Recognition
}

\author{
  Hao Li$^{1}$, Mingrui Zheng$^{2}$, Yasuyuki Tahara$^{1}$, Yuichi Sei$^{1}$ \\
  $^{1}$Department of Informatics, Graduate School of Informatics and Engineering \\
  The University of Electro-Communications, Tokyo, Japan \\
  $^{2}$Graduate School of Information Science and Technology \\
  Hokkaido University, Sapporo, Japan \\
  \texttt{r2540015@gl.cc.uec.ac.jp, tahara@uec.ac.jp, seiuny@uec.ac.jp} \\
  \texttt{zmr1584383567@stl.ssi.ist.hokudai.ac.jp} \\
  \texttt{ORCID(Hao Li): 0009-0004-2623-286X} \\
  \texttt{ORCID(Mingrui Zheng): 0009-0008-1299-1821} \\
  \texttt{ORCID(Yasuyuki Tahara): 0000-0002-1939-4455} \\
  \texttt{ORCID(Yuichi Sei): 0000-0002-2552-6717}
}

\begin{document}
\maketitle

\begin{abstract}
Recent studies on sensor-language alignment have shown that two-stage frameworks can improve the semantic modeling ability of wearable-sensor human activity recognition (HAR), where SensorLLM-style methods first perform motion-to-language alignment and then fine-tune the model for downstream tasks \cite{li2025sensorllm}. However, our experiments reveal a consistent failure mode when the Stage 2 backbone is compressed to a compact model such as TinyLlama: recognition of dynamic activities remains relatively strong, while the discrimination of low-motion static classes such as standing, sitting, and lying degrades substantially. To address this issue, we propose a gravity-aware hierarchical routing head as a lightweight post-alignment adaptation built on top of an already aligned model, rather than a new large-scale pretraining framework. The method uses the per-channel mean and std from the Chronos tokenizer state to extract statistical cues related to posture and gravity direction, and adaptively combines a static expert and a full expert through soft routing, together with a load-balancing loss for stable training. On the MHealth dataset \cite{banos2014mhealth}, this design significantly improves macro-F1 with minimal parameter overhead, and the gains are concentrated mainly on static classes while preserving strong performance on dynamic activities. As a first arXiv disclosure, the current paper reports results on a single dataset only, with the goal of highlighting the core method and laying the groundwork for broader evaluation in future work.
\end{abstract}

\keywords{human activity recognition \and wearable sensors \and sensor-language alignment \and lightweight large language models \and hierarchical routing \and static activity recognition}

\section{Introduction}

Human Activity Recognition (HAR) is one of the core tasks in wearable computing, mobile health, and ambient intelligence. Its goal is to infer the current activity state of a person from time-series signals collected by sensors such as accelerometers and gyroscopes, for example walking, running, sitting, or lying \cite{lima2019haroverview}. Traditional HAR methods are mostly built on temporal models such as convolutional networks, recurrent networks, or Transformers, and perform classification by learning discriminative representations from raw multichannel sensor data \cite{ordonez2016deepconvlstm}. In recent years, as large language models (LLMs) have demonstrated strong capabilities in general reasoning, knowledge integration, and cross-modal expression, researchers have begun to explore how to align numerical time-series signals with natural language, so that models can not only "recognize activities" but also explain sensor patterns in a way that is closer to human understanding \cite{li2025sensorllm}.

SensorLLM is one of the representative works along this direction. Its core idea is to align sensor time series with natural-language descriptions in two stages: first, in stage 1, it establishes sensor-language alignment through question-answer style data constructed from single-channel sensor segments together with trend text and statistical summaries; then, in stage 2, it uses the aligned model for downstream HAR classification in the multichannel setting \cite{li2025sensorllm}. The importance of this design lies in the fact that it no longer treats HAR merely as a supervised classification problem over a closed label space, but instead attempts to build a bridge between "sensor patterns" and "linguistic semantics" within an LLM. Such a framework is clearly appealing for complex sensor inputs, interpretable prediction, and broader future tasks in time-series understanding.

However, the original SensorLLM relies on a relatively large language-model backbone, which imposes a high cost in deployment, training resources, and experimental iteration speed. For many practical research and application scenarios, especially rapid prototyping or resource-constrained environments, a natural question is whether its core capabilities can be reproduced with a smaller language model. Motivated by this question, we replace the original larger LLM with TinyLlama-1.1B \cite{zhang2024tinyllama} and build a lightweight SensorLLM baseline. Our experiments show that this lightweight substitution does not degrade performance uniformly across all classes. Instead, we observe a fairly stable failure mode: for dynamic activities such as walking, jogging, and running, the small model still maintains relatively acceptable recognition performance; but for static activities such as standing, sitting, and lying, the inter-class discrimination ability drops substantially. This phenomenon suggests that the bottleneck of the small model does not arise solely from insufficient overall capacity, but more likely from inadequate representation of certain fine-grained, low-energy, posture-sensitive classes.

This observation further leads to the central question of this paper: when a lightweight SensorLLM suffers from confusion among static activities in stage-2 classification, can this weakness be compensated not by enlarging the LLM itself, but through a lighter and more structurally targeted classification head? Our intuition is that the discriminative basis of static and dynamic activities is not the same. Dynamic activities are usually accompanied by larger amplitude variation, stronger temporal fluctuation, and more distinctive motion patterns, and are therefore relatively easy to distinguish through differences in "energy" within multichannel sensor signals. The difficulty among static activities, by contrast, is precisely that they all lack strong motion, while differing in subtle yet crucial ways in posture, sensor orientation, and gravity projection patterns. For example, standing, sitting, and lying may all have low overall motion intensity, but often differ in gravity distribution across axes. For such classes, a single shared linear classification head may not be the most suitable choice.

Based on this motivation, we propose a stage-2 classification enhancement module for lightweight SensorLLM: a gravity-aware hierarchical routing head. This method does not change the overall two-stage SensorLLM framework, but instead focuses on the downstream classification head, a local yet critical component. Specifically, we first construct a hierarchical routing structure that allows the model to perform soft allocation between two experts according to sensor priors: a static expert specialized for fine-grained discrimination of static activities, and a full expert responsible for all-class prediction. Compared with forcing a single classification head to handle all classes simultaneously, such a division of labor is better aligned with the practical needs of the lightweight setting, where limited discriminative capacity should be concentrated on the most difficult subproblem.

On top of this hierarchical structure, we further design two types of routing signals. The first is energy-aware routing, which constructs routing features based on the standard deviation of each sensor channel and their aggregated statistics, in order to coarsely determine whether a sample is closer to "static" or "dynamic." The second, which is the main focus of this paper, is gravity-aware routing. The key point here is not to claim that gravity is used in HAR for the first time, since the use of gravity-direction information for posture recognition has already been well studied \cite{vahaypya2018posture}. Rather, the contribution of this paper is to explicitly inject this physical prior into a lightweight stage-2 routing head in a SensorLLM-compatible and extremely low-overhead manner. More concretely, we use the per-channel mean and std already available in the tokenizer state as a compact representation of posture-related gravity information and motion-intensity information, and use these features both for routing decisions and for the discrimination process of the static expert. The advantage of doing so is that it avoids redesigning the entire temporal backbone and does not rely on complicated additional preprocessing, but instead directly reuses statistical information already produced by the existing pipeline.

Formally, the proposed method has three characteristics. First, it is lightweight: the added modules operate only near the classification head, and the parameter increase is far smaller than the cost of replacing the model with a larger LLM. Second, it is plug-and-play: the method does not alter the stage-1 alignment process of SensorLLM and does not require reconstruction of the entire input representation. Third, it is driven by physical priors: rather than relying entirely on LLM hidden states for unified discrimination, this paper explicitly uses energy and gravity cues in the sensor data, decomposing the problem of "why static classes are hard to separate" into a more structured combination of routing and specialized classification.

This paper is currently validated mainly on the MHealth dataset. More precisely, it is a focused technical report centered on a single dataset and a single bottleneck, rather than a work that attempts to make comprehensive claims of cross-dataset generality. We adopt this framing because our goal is first to clearly identify a key failure mode of lightweight SensorLLM and to verify a structurally simple, intuitively motivated, and experimentally reproducible improvement direction, rather than to extend prematurely to broader conclusions that have not yet been sufficiently validated.

The main contributions of this paper are as follows.
First, we identify and systematically characterize a specific failure mode of lightweight SensorLLM for HAR: small models maintain relatively good performance on dynamic activities but degrade substantially in discriminating static activities.
Second, we propose a lightweight hierarchical routing classification head that decomposes stage-2 HAR classification into two subtasks: static-sensitive fine-grained discrimination and full-class prediction.
Third, we propose a gravity-aware routing variant that explicitly injects the mean/std in the tokenizer state as posture-related physical priors into the routing and static classification process of lightweight SensorLLM.
Fourth, our preliminary experiments on MHealth show that the proposed method can significantly improve static-class recognition with minimal parameter overhead, providing a promising direction for future studies on more datasets and device settings.

\section{Related Work}

\subsection{Sensor-Language Alignment and Wearable Foundation Models}

\textbf{Directly related work.} SensorLLM is the work most directly related to this paper, as it aligns motion sensors with language through a two-stage framework and performs downstream task-aware tuning for HAR on top of that alignment \cite{li2025sensorllm}. Our work follows this overall problem setting, but focuses not on redesigning the alignment pipeline, but on modifying the Stage 2 classifier under a compact-backbone setting \cite{li2025sensorllm}.

\textbf{Partially related work.} SensorLM extends sensor-language modeling to larger-scale wearable scenarios by constructing large-scale paired sensor-text data, with a focus on demonstrating capabilities such as zero-shot recognition, few-shot learning, cross-modal retrieval, and sensor captioning \cite{zhang2025sensorlm}. SLIP is also an important neighboring work in this direction, focusing on learning transferable sensor representations through language-informed pretraining and supporting tasks such as heterogeneous sensor configurations, captioning, and question answering \cite{chen2026slip}. Time2Lang is partially related to our work because it studies how to directly bridge time-series foundation models and LLMs in health sensing tasks without relying on explicit text prompts, but its core problem is representation mapping rather than HAR classification-head design \cite{pillai2025time2lang}. ChatTS is also partially related because it builds multimodal large models for more general time-series understanding and reasoning, but its training emphasis is on generic time-series reasoning with synthetic aligned data rather than a two-stage lightweight classification setting for wearable HAR \cite{xie2024chatts}.

\textbf{Broad background work.} Scaling Wearable Foundation Models provides important background for large-scale wearable representation learning and shows that foundation-model scaling can improve downstream tasks including activity recognition, but it does not study the design of sensor-language task heads in compact HAR classifiers \cite{narayanswamy2024scalingwearable}. PH-LLM is also an important background reference because it demonstrates LLM capabilities for understanding and reasoning over wearable and personal health time series, but its primary goals are health insight generation, recommendation, and outcome prediction rather than structured adaptation of HAR classification heads \cite{cosentino2024phllm}.

\subsection{HAR and the Challenge of Static Activity Recognition}

Wearable-sensor-based HAR has been extensively studied, and one persistent challenge is that low-motion postural classes are often harder to distinguish finely than dynamic activities, even though they appear simpler on the surface \cite{lima2019haroverview}. Specifically, standing, sitting, and lying often share weak short-term dynamics and low motion energy, making them more prone to confusion when a model relies mainly on motion intensity rather than posture-sensitive cues \cite{vahaypya2018posture,lima2019haroverview}. This issue does not disappear automatically even with stronger representation-learning models, because stronger semantic alignment does not necessarily translate into improved fine-grained discrimination of static classes by compact downstream classifiers \cite{li2025sensorllm}.

Among the works discussed above, SensorLLM is the closest to the main HAR task setting considered here, but it does not explicitly redesign the classifier backend for static-activity discrimination under a compact-backbone setting \cite{li2025sensorllm}. Although SensorLM and SLIP cover activity-related recognition or transfer scenarios, their main contributions remain centered on large-scale pretraining and generalization rather than correcting posture confusion in a lightweight HAR head \cite{zhang2025sensorlm,chen2026slip}. Time2Lang and PH-LLM are further from this problem, because they are primarily oriented toward health sensing and wearable reasoning rather than dedicated modeling of static-class confusion in benchmark HAR \cite{pillai2025time2lang,cosentino2024phllm}.

\subsection{Routing / Mixture-of-Experts / Adaptive Heads}

Routing, mixture-of-experts, and adaptive heads provide a general design principle: when different input patterns correspond to different discriminative requirements, model capacity can be allocated through specialized decision paths rather than relying on a single shared classification head to predict all samples \cite{shazeer2017moe,fedus2022switch}. This idea is particularly meaningful for wearable HAR, because static and dynamic activities differ substantially in signal structure, and a fully shared decision head may not be the most suitable design \cite{shazeer2017moe,fedus2022switch}.

However, the sensor-language works most relevant to this paper still focus mainly on alignment, pretraining, scaling, or more general multimodal reasoning, rather than hierarchical classifier-side routing in compact HAR models. SensorLLM emphasizes two-stage sensor-language alignment for HAR \cite{li2025sensorllm}, SensorLM and SLIP emphasize large-scale sensor-language pretraining and transfer \cite{zhang2025sensorlm,chen2026slip}, Time2Lang focuses on bridging TFMs and LLMs \cite{pillai2025time2lang}, ChatTS focuses on more general time-series understanding and reasoning \cite{xie2024chatts}, while Scaling Wearable Foundation Models and PH-LLM provide broader background on wearable foundation models and wearable LLMs \cite{narayanswamy2024scalingwearable,cosentino2024phllm}. Therefore, to the best of our knowledge, existing work has not studied a gravity-aware hierarchical routing backend for compact SensorLLM-based HAR.

\section{Method}

\subsection{Problem Setup}

This paper studies wearable-sensor human activity recognition (HAR) under the two-stage SensorLLM framework \cite{li2025sensorllm}. Let the training set be
\begin{equation}
\mathcal{D}=\{(\mathbf{x}_i,y_i)\}_{i=1}^{N},
\end{equation}
where $\mathbf{x}_i \in \mathbb{R}^{C \times T}$ denotes a multivariate sensor sequence with $C$ channels and length $T$, and $y_i \in \{1,\dots,K\}$ denotes the activity label. In the current first arXiv version, we instantiate the method on MHealth, where $K=12$, and the static subset corresponds to standing, sitting, and lying.

\begin{figure*}[t]
    \centering
    \includegraphics[width=\textwidth]{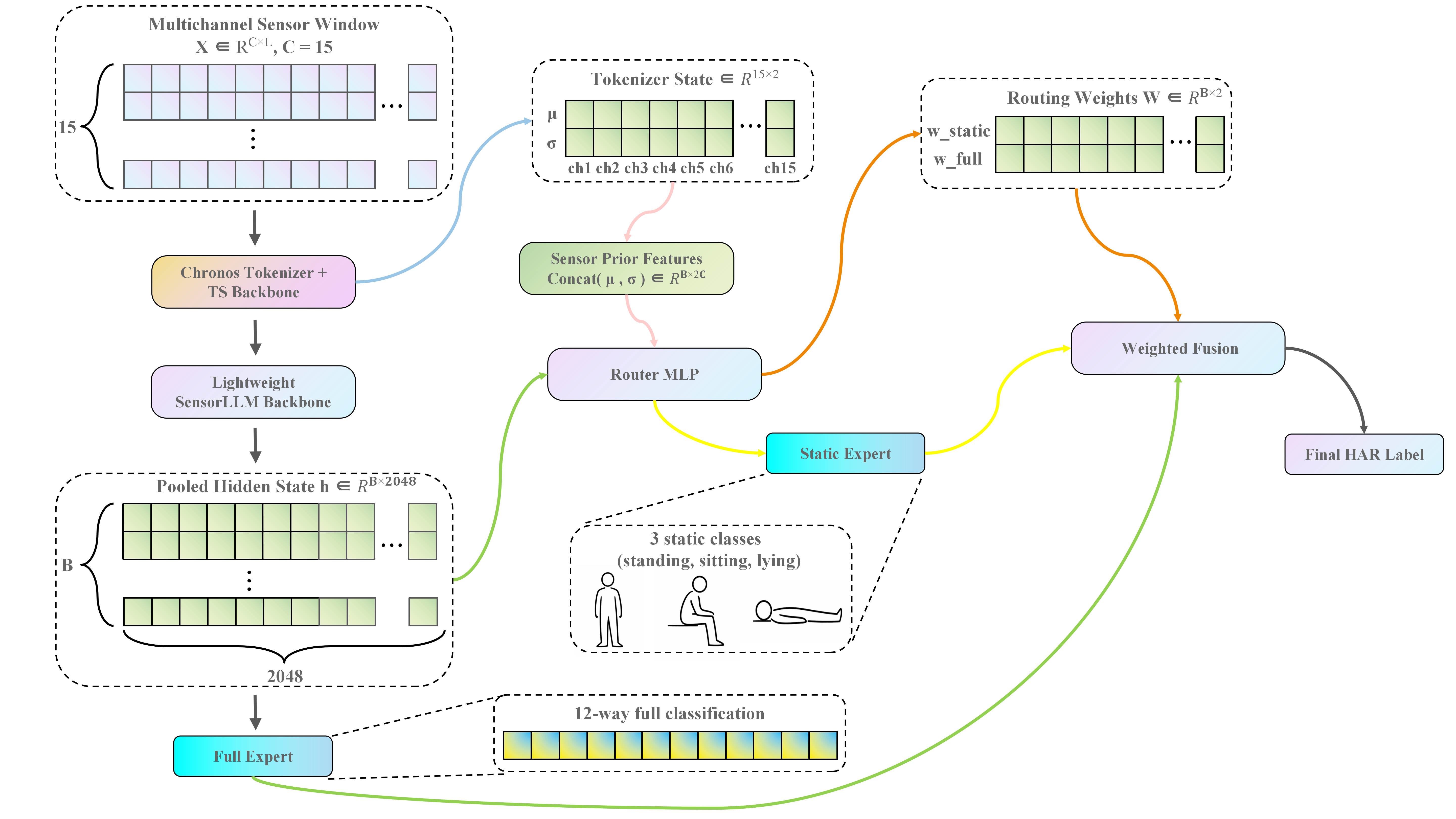}
    \caption{
    Overview of the proposed gravity-aware hierarchical routing head. The input multichannel sensor window is first processed by the Chronos tokenizer / time-series backbone, and then fed into the lightweight SensorLLM stage-2 backbone to obtain the pooled hidden representation $\mathbf{h}$. Meanwhile, the Chronos tokenizer state provides the mean and standard deviation of each channel, which together form sensor prior features that characterize statistical information related to gravity direction and energy. The router MLP then generates sample-wise routing weights from these priors, and uses them to combine two experts: a static expert specialized for standing, sitting, and lying, and a full expert responsible for prediction over all classes. The figure also marks the energy-aware variant used in the ablation study, which performs routing using standard-deviation statistics only.}
    \label{fig:method_overview}
\end{figure*}

Following the setting of SensorLLM \cite{li2025sensorllm}, Stage 1 is responsible for sensor-language alignment and is kept completely unchanged in this paper. Our modification is limited to Stage 2, where sequence classification is performed on top of the aligned backbone. Figure~\ref{fig:method_overview} illustrates the overall pipeline. The most important point is that we preserve the already aligned SensorLLM backbone and replace only the lightweight stage-2 classification head.

For each input sample, the Stage 2 backbone outputs a hidden-state sequence
\begin{equation}
\mathbf{H}_i = F_{\theta}(\mathbf{x}_i) \in \mathbb{R}^{L_i \times d},
\end{equation}
where $L_i$ denotes the output token length and $d$ denotes the hidden dimension. We use the hidden state of the last valid token as the pooled representation:
\begin{equation}
\mathbf{h}_i = \mathrm{Pool}(\mathbf{H}_i) = \mathbf{H}_i[\ell_i], 
\qquad
\ell_i = \max \{t \mid m_{i,t}=1\},
\end{equation}
where $m_{i,t}$ is the attention mask.

In addition to $\mathbf{H}_i$, the Chronos tokenizer also provides a tokenizer state for each channel, which contains the channel mean and standard deviation. We denote them by $\boldsymbol{\mu}_i,\boldsymbol{\sigma}_i \in \mathbb{R}^{C}$, and define the sensor-statistics vector as
\begin{equation}
\mathbf{s}_i = [\boldsymbol{\mu}_i;\boldsymbol{\sigma}_i] \in \mathbb{R}^{2C}.
\end{equation}
As shown in Figure~\ref{fig:method_overview}, $\boldsymbol{\mu}_i$ mainly carries offset information related to posture and gravity direction, while $\boldsymbol{\sigma}_i$ reflects activity intensity. Together, these statistics form the sensor priors used by the routing module in this work.

\subsection{Baseline: Stage-2 SensorLLM Classifier}

The original Stage-2 SensorLLM classifier places a shared linear classification head on top of the pooled hidden state \cite{li2025sensorllm}. Formally, the baseline logits are
\begin{equation}
\mathbf{z}_i^{\mathrm{base}} = W_{\mathrm{cls}} \mathbf{h}_i,
\end{equation}
where $W_{\mathrm{cls}} \in \mathbb{R}^{K \times d}$. The final prediction is
\begin{equation}
\hat{y}_i^{\mathrm{base}} = \arg\max_{k \in \{1,\dots,K\}} z_{i,k}^{\mathrm{base}}.
\end{equation}

This design is simple and effective, but it requires the same decision head to handle both high-motion dynamic activities and low-motion static activities. As shown in Figure~\ref{fig:method_overview}, our method keeps the backbone unchanged and replaces only this lightweight stage-2 classification head.

\subsection{Gravity-Aware Hierarchical Routing}

Let the static label set be $\mathcal{Y}_{s} \subset \{1,\dots,K\}$. In the current MHealth setting, $\mathcal{Y}_{s}$ corresponds to three static classes, so $K_s = |\mathcal{Y}_{s}| = 3$. On top of the pooled representation, we add a lightweight two-expert hierarchical classification head.

\paragraph{Router.}
The router takes the sensor-statistics vector $\mathbf{s}_i$ as input and outputs sample-wise routing weights:
\begin{equation}
\mathbf{r}_i = \mathrm{softmax}\!\left(\frac{g_{\phi}(\mathbf{s}_i)}{\tau}\right)
= [\alpha_i,\beta_i],
\qquad
\alpha_i + \beta_i = 1,
\end{equation}
where $g_{\phi}$ denotes a small MLP router and $\tau$ is the routing temperature. In the current implementation, the router has the architecture $2C \rightarrow 64 \rightarrow 32 \rightarrow 2$. Figure~\ref{fig:method_overview} clearly shows that the router depends only on the sensor priors extracted from the tokenizer state, rather than operating directly on the full hidden sequence.

\paragraph{Static expert.}
The static expert receives both the pooled hidden state and the sensor-statistics vector:
\begin{equation}
\tilde{\mathbf{z}}_i^{\mathrm{stat}}
= E_{\mathrm{stat}}([\mathbf{h}_i;\mathbf{s}_i])
\in \mathbb{R}^{K_s}.
\end{equation}
Since this expert predicts only the static subset, we expand it back to the full label space using a zero-filling projection operator $\Pi_{\mathcal{Y}_{s}}$:
\begin{equation}
\mathbf{z}_i^{\mathrm{stat}}
=
\Pi_{\mathcal{Y}_{s}}(\tilde{\mathbf{z}}_i^{\mathrm{stat}})
\in \mathbb{R}^{K}.
\end{equation}
After expansion, the values at non-static class positions are set to 0. In implementation, this expert is an MLP with architecture $(d+2C) \rightarrow 256 \rightarrow 64 \rightarrow K_s$.

\paragraph{Full expert.}
The full expert behaves similarly to the baseline and predicts all classes using only the pooled hidden state:
\begin{equation}
\mathbf{z}_i^{\mathrm{full}} = E_{\mathrm{full}}(\mathbf{h}_i) = W_{\mathrm{full}}\mathbf{h}_i,
\end{equation}
where $W_{\mathrm{full}} \in \mathbb{R}^{K \times d}$.

\paragraph{Soft logit fusion.}
The final output logits are obtained by weighted fusion of the two experts using sample-wise routing weights:
\begin{equation}
\mathbf{z}_i
=
\alpha_i \mathbf{z}_i^{\mathrm{stat}}
+
\beta_i \mathbf{z}_i^{\mathrm{full}}.
\end{equation}
As shown in Figure~\ref{fig:method_overview}, the full expert is responsible for prediction over all classes, while the static expert specializes in standing, sitting, and lying. Soft routing keeps the entire model end-to-end differentiable while avoiding the instability introduced by hard routing in early training.

\paragraph{Intuition.}
The core intuition behind our method is that static activities are low-energy, but not information-free. If the classifier mainly relies on short-term motion magnitude, then standing, sitting, and lying are easily confused; however, from the perspective of inertial sensors, these three activities still preserve stable differences in statistics related to gravity direction and posture. Channel means encode direction-related offset information, while channel standard deviations characterize residual motion intensity. Our method uses these signals in two complementary ways: on the one hand, the router uses $\mathbf{s}_i$ to determine to what extent a sample should be sent to a static-specialized path; on the other hand, the static expert directly receives the same statistics to strengthen posture-sensitive discrimination. For this reason, the proposed structure is particularly suitable for fine-grained discrimination among static classes, while requiring no modification to the already aligned backbone.

\subsection{Training Objective}

All routing experiments in this paper are trained with class-weighted cross-entropy, where class weights are derived from the normalized inverse of label frequencies in the training set. Let the empirical frequency of class $k$ in the training set be $n_k$. We define
\begin{equation}
\omega_k = \frac{n_k^{-1}}{\sum_{j=1}^{K} n_j^{-1}}.
\end{equation}

The corresponding weighted classification loss is
\begin{equation}
\mathcal{L}_{\mathrm{wce}}
=
-\frac{1}{N}
\sum_{i=1}^{N}
\omega_{y_i}
\log
\frac{\exp(z_{i,y_i})}
{\sum_{k=1}^{K}\exp(z_{i,k})}.
\end{equation}

To avoid router collapse, we additionally introduce a load-balancing term that encourages the average routing mass to be distributed more evenly across experts. Let
\begin{equation}
\bar{\mathbf{r}} = \frac{1}{N}\sum_{i=1}^{N}\mathbf{r}_i.
\end{equation}
The load-balancing loss is then written as
\begin{equation}
\mathcal{L}_{\mathrm{lb}}
=
\lambda
\left\|
\bar{\mathbf{r}} - \frac{1}{2}\mathbf{1}
\right\|_2^2,
\end{equation}
where $\lambda$ is the balancing coefficient. In the current implementation, $\lambda=0.1$.

The final training objective is
\begin{equation}
\mathcal{L}
=
\mathcal{L}_{\mathrm{wce}}
+
\mathcal{L}_{\mathrm{lb}}.
\end{equation}
When routing is not used, $\mathcal{L}_{\mathrm{lb}}$ is removed.

\subsection{Ablation Variants}

To analyze the contribution of each component, we design four ablation variants.

\paragraph{Ablation A: Feature-Augmented Head.}
This variant removes routing and directly injects the sensor statistics into a single MLP classification head:
\begin{equation}
\mathbf{z}_i^{A} = E_{A}([\mathbf{h}_i;\mathbf{s}_i]).
\end{equation}
It is used to test whether feature injection alone is sufficient to explain the performance improvement.

\paragraph{Ablation B: Large MLP Head.}
This variant removes both routing and sensor-statistics injection, and compares performance by increasing only the capacity of the classification head:
\begin{equation}
\mathbf{z}_i^{B} = E_{B}(\mathbf{h}_i).
\end{equation}
It is used to test whether the gain comes solely from increased parameter count.

\paragraph{Ablation C: Energy-Only Routing.}
This variant preserves the routing structure, but uses only the standard-deviation vector, i.e., energy information, without using channel means:
\begin{equation}
\mathbf{r}_i^{C}
=
\mathrm{softmax}\!\left(\frac{g_{\phi}^{C}(\boldsymbol{\sigma}_i)}{\tau}\right),
\end{equation}
\begin{equation}
\mathbf{z}_i^{C}
=
\alpha_i^{C}
\Pi_{\mathcal{Y}_{s}}\!\left(
E_{\mathrm{stat}}^{C}([\mathbf{h}_i;\boldsymbol{\sigma}_i])
\right)
+
\beta_i^{C} E_{\mathrm{full}}^{C}(\mathbf{h}_i).
\end{equation}
This version is the energy-aware variant marked in Figure~\ref{fig:method_overview}. It is used to test whether the mean information related to gravity and posture is a key factor.

\paragraph{Ablation D: Routing Only Without Feature Injection into the Static Expert.}
This variant preserves the gravity-aware router, but the static expert receives only the pooled hidden state:
\begin{equation}
\mathbf{z}_i^{D}
=
\alpha_i^{D}
\Pi_{\mathcal{Y}_{s}}\!\left(
E_{\mathrm{stat}}^{D}(\mathbf{h}_i)
\right)
+
\beta_i^{D} E_{\mathrm{full}}^{D}(\mathbf{h}_i).
\end{equation}
It is used to test whether the routing structure alone is already sufficient to provide the main gain, without directly injecting sensor statistics into the static expert.

\section{Experimental Setup}

\subsection{Dataset and Task}

We conduct experiments on the MHealth dataset \cite{banos2014mhealth,banos2014mhealthdroid}. MHealth is a classic benchmark dataset for wearable-sensor human activity recognition (HAR). In the current study, we use the MHealth stage-2 classification split generated by our project pipeline and formulate the task as a 12-class closed-set classification problem based on multichannel sensor sequences. Each sample contains 15 sensor channels, and the evaluation target is standard HAR classification performance.

To keep this first arXiv version as focused and easy to verify as possible, we intentionally limit the experimental scope to a single dataset. At this stage, the goal is to disclose the method itself as early as possible, rather than claim that broad cross-dataset generalization has already been established. On MHealth, our method primarily targets the confusion among the three static classes, standing, sitting, and lying, while aiming to preserve strong performance on the remaining dynamic activities.


\begin{table}[t]
\centering
\footnotesize
\setlength{\tabcolsep}{4pt}
\renewcommand{\arraystretch}{1.08}
\caption{Dataset and task configuration used in the current single-dataset first report.}
\label{tab:dataset_task}
\begin{tabularx}{\linewidth}{l Y c c Y Y l}
\toprule
Dataset & Task & \#Classes & \#Channels & Static subset & Split used & Main metric \\
\midrule
MHealth \cite{banos2014mhealth} & wearable HAR & 12 & 15 & standing / sitting / lying & processed stage-2 train / eval split & macro-F1 \\
\bottomrule
\end{tabularx}
\end{table}

\subsection{Implementation Details}

Our experiments follow a SensorLLM-style two-stage setup \cite{li2025sensorllm}. Stage 1 is used for sensor-language alignment and is kept fixed throughout this paper; Stage 2 is initialized from the aligned stage-1 checkpoint, using TinyLlama-1.1B \cite{zhang2024tinyllama} as the compact language backbone and Chronos-T5-Large \cite{ansari2024chronos} as the time-series encoder. Sensor sequences are processed using the Chronos-style \texttt{StanNormalizeUniformBins} tokenizer \cite{ansari2024chronos}, while the text-side input in Stage 2 follows the summary-plus-question prompt format generated by the project pipeline.

Unless otherwise stated, all compared methods share exactly the same training configuration. We train for 10 epochs, with a per-device batch size of 8 for both training and evaluation. The learning rate is set to $2\times 10^{-3}$, the learning-rate scheduler is cosine, the warmup ratio is 0.03, weight decay is 0, and the maximum norm for gradient clipping is 1.0. FP16 and gradient checkpointing are enabled during training. The model is saved and evaluated every 50 steps, and the best checkpoint is selected according to evaluation macro-F1. In the current implementation, both the aligned LLM backbone and the Chronos encoder are frozen, and only the stage-2 classification head, or the corresponding routing head in our method, is trained. For routing-based methods, the router temperature is fixed at 1.0.

Following the repository training pipeline, we use class-weighted cross-entropy, where class weights are computed from the normalized inverse of label frequencies in the training set. All main experiments are repeated over five random seeds: 42, 123, 456, 789, and 2024. To ensure fair comparison, the baseline, our method, and all ablation variants use exactly the same data split, optimization hyperparameters, random-seed set, and model-selection protocol.

One additional detail related to transparency should be clarified. The current first-release pipeline provides train and evaluation splits, but does not include a separate validation split. Therefore, the best checkpoint is selected according to macro-F1 on the evaluation split, and the final results are also reported on the same evaluation split. To maintain comparability across all methods, we keep this setting unchanged and position the current version as a first arXiv disclosure focused primarily on the method, rather than a finalized benchmark study.


\begin{table}[t]
\centering
\small
\setlength{\tabcolsep}{6pt}
\caption{Reproducible experimental configuration used in the current study.}
\label{tab:implementation_details}
\begin{tabular}{l l}
\toprule
Component & Setting \\
\midrule
Stage-1 init checkpoint & aligned SensorLLM-style checkpoint \cite{li2025sensorllm} \\
LLM backbone & TinyLlama-1.1B \cite{zhang2024tinyllama} \\
TS encoder & Chronos-T5-Large \cite{ansari2024chronos} \\
Tokenization & StanNormalizeUniformBins \cite{ansari2024chronos} \\
Epochs & 10 \\
Train / eval batch size & 8 / 8 \\
Learning rate & $2 \times 10^{-3}$ \\
Scheduler & cosine \\
Warmup ratio & 0.03 \\
Weight decay & 0.0 \\
Precision & FP16 \\
Gradient checkpointing & yes \\
Checkpoint selection & best eval macro-F1 \\
Seeds & 42, 123, 456, 789, 2024 \\
\bottomrule
\end{tabular}
\end{table}

\subsection{Evaluation Metrics}

Our primary metric is macro-F1, i.e., the unweighted average of class-wise F1 scores over the 12 classes. We choose macro-F1 as the main selection and reporting metric because the proposed method is designed to address a structured failure mode concentrated in a small subset of classes, and macro-F1 reflects such class-level improvements more effectively than overall accuracy. In addition to macro-F1, we also report accuracy and micro-F1 as auxiliary aggregate metrics, and further examine per-class F1 scores to analyze where the gains primarily come from.

All main results are reported as mean and standard deviation over five random seeds. In addition to aggregate metrics, we place particular emphasis on the per-class F1 of the three static activities, because these are the primary targets of the proposed routing mechanism. Under the current setting, multi-seed evaluation is necessary because the compact stage-2 classifier exhibits noticeable variation across random seeds, especially in the fine-grained discrimination of static classes.


\subsection{Baseline and Compared Variants}

We compare the following models under exactly the same experimental protocol. \textbf{Baseline} refers to a compact SensorLLM-style stage-2 classifier \cite{li2025sensorllm}, which pools the final hidden state and applies a shared linear classification head. This is the primary reference point for evaluating whether routing can improve compact post-alignment HAR. \textbf{Ours} replaces this shared classification head with the proposed gravity-aware hierarchical routing module.

In addition, we evaluate four ablation variants. \textbf{Ablation A} removes routing and uses a feature-augmented MLP head that directly receives the pooled hidden state and sensor statistics. \textbf{Ablation B} removes both routing and sensor-statistics injection, and instead uses a larger MLP head to control for the effect of parameter count. \textbf{Ablation C} preserves the routing structure but uses only per-channel standard deviations, removing the mean statistics related to gravity and posture. \textbf{Ablation D} preserves gravity-aware routing but no longer injects sensor statistics directly into the static expert. Through these variants, we can separately analyze the contributions of the routing structure, gravity-sensitive statistics, and feature injection itself.

\begin{figure*}[t]
    \centering
    \includegraphics[width=\textwidth]{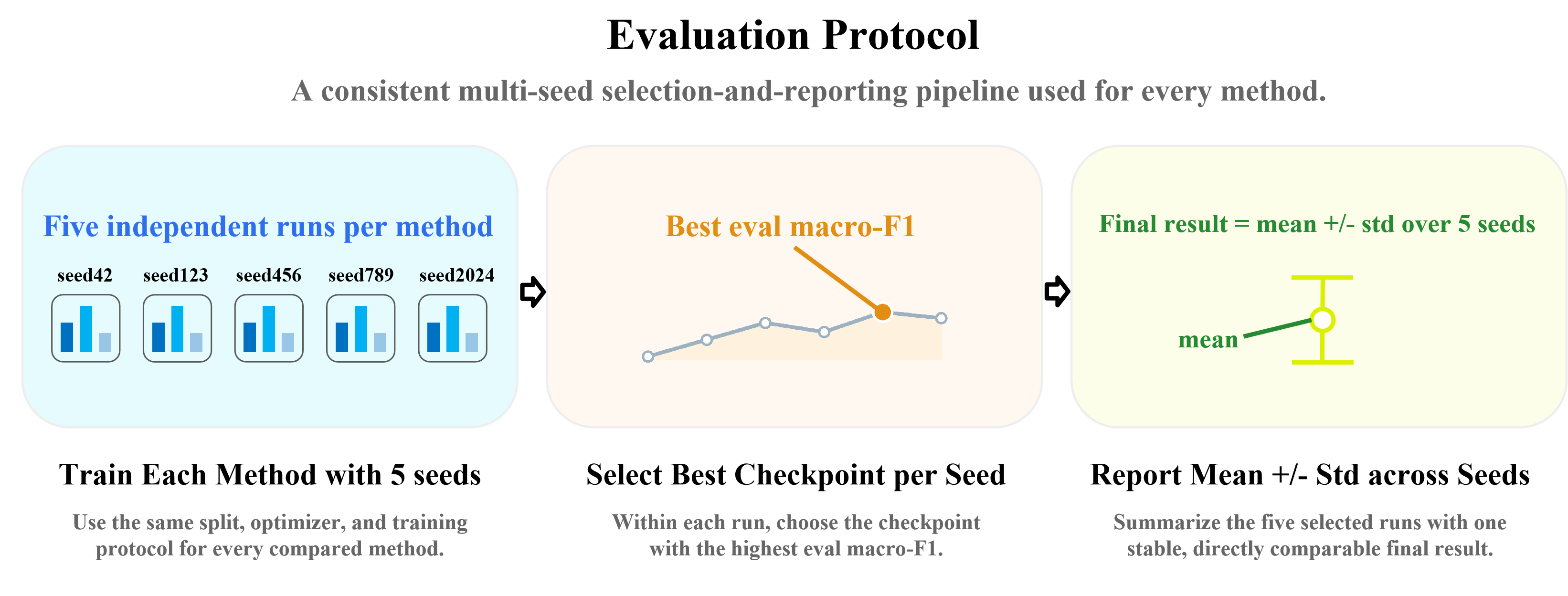}
    \caption{
    Illustration of the unified evaluation protocol used for all methods in this paper. Each method is trained with five random seeds under the same data split and optimization settings; for each seed, the best checkpoint is selected according to eval macro-F1, and the final result is reported as the mean and standard deviation (mean $\pm$ std) across seeds. This protocol ensures consistency in the model-selection procedure and makes comparisons across methods statistically more transparent.}
    \label{fig:eval_protocol}
\end{figure*}

\begin{table*}[t]
\centering
\footnotesize
\setlength{\tabcolsep}{4pt}
\renewcommand{\arraystretch}{1.10}
\caption{Structural comparison of the baseline, ablation variants, and the proposed method. All variants reuse the same aligned Stage-1 checkpoint.}
\label{tab:method_variants}
\begin{tabularx}{\textwidth}{l Y l Y c Y}
\toprule
Method & Stage-2 head & Router input & Static expert input & Multi-seed & Purpose \\
\midrule
Baseline   & pooled hidden + linear head        & --        & --                                  & yes & compact SensorLLM reference \\
Ablation A & feature-augmented MLP              & --        & $[\mathbf{h}_i;\mathbf{s}_i]$       & yes & test feature injection only \\
Ablation B & larger MLP                         & --        & $\mathbf{h}_i$                      & yes & test parameter count only \\
Ablation C & routing head                       & std only  & $[\mathbf{h}_i;\boldsymbol{\sigma}_i]$ & yes & test energy-only routing \\
Ablation D & routing head                       & mean+std  & $\mathbf{h}_i$                      & yes & test routing without feature injection \\
Ours       & gravity-aware hierarchical routing & mean+std  & $[\mathbf{h}_i;\mathbf{s}_i]$       & yes & full method \\
\bottomrule
\end{tabularx}
\end{table*}

\section{Results}

\subsection{Main Results}

Table~\ref{tab:main_results} summarizes three core reference points on MHealth: the published SensorLLM result reported in the original paper \cite{li2025sensorllm}, our TinyLlama-based compact baseline, and the proposed gravity-aware hierarchical routing model. The original paper reports $89.4 \pm 3.85$ macro-F1 and $89.0 \pm 3.54$ accuracy on MHealth \cite{li2025sensorllm}. Under our compact Stage 2 setting, the TinyLlama baseline reaches $89.97 \pm 2.19$ macro-F1 and $89.28 \pm 2.39$ accuracy, showing that the aligned SensorLLM-style pipeline remains competitive even after substantially reducing the classifier backbone.

When we replace the shared linear classification head with gravity-aware hierarchical routing, performance further improves to $96.44 \pm 0.82$ macro-F1 and $96.24 \pm 0.87$ accuracy, corresponding to absolute gains of $+6.47$ and $+6.96$ points over the compact baseline, respectively. This is also the central empirical conclusion of the paper: the main bottleneck of the current compact baseline is not inadequate recognition of dynamic activities as a whole, but rather a structured decision failure mode that can be effectively corrected by a lightweight routing head.

At the same time, we do not interpret the comparison with the original SensorLLM paper as a strictly controlled head-to-head conclusion. The original paper uses a different backbone scale and different training details, whereas the current version of our work is intentionally focused on a first method disclosure under a single-dataset, compact-backbone setting. Therefore, the published SensorLLM result mainly serves as background context; the truly controlled and most important comparison in this paper is between the \emph{compact baseline and gravity-aware routing} under exactly the same local protocol \cite{li2025sensorllm}.


\begin{table}[t]
\centering
\footnotesize
\setlength{\tabcolsep}{4pt}
\renewcommand{\arraystretch}{1.08}
\caption{Main results on MHealth. The published SensorLLM result is included for context only and is not a strictly controlled comparison because the backbone and training setup differ from our local protocol.}
\label{tab:main_results}
\begin{tabularx}{\linewidth}{l l X c c l}
\toprule
Method & Backbone & Stage-2 head & \shortstack{Macro-F1\\(\%)} & \shortstack{Accuracy\\(\%)} & Note \\
\midrule
SensorLLM (paper) \cite{li2025sensorllm} & paper backbone & original Stage-2 head & 89.4 $\pm$ 3.85 & 89.0 $\pm$ 3.54 & context only \\
\midrule
Compact baseline & TinyLlama-1.1B & pooled hidden + linear head & 89.97 $\pm$ 2.19 & 89.28 $\pm$ 2.39 & 5 seeds \\
Gravity routing & TinyLlama-1.1B & gravity-aware hierarchical routing & \textbf{96.44 $\pm$ 0.82} & \textbf{96.24 $\pm$ 0.87} & 5 seeds \\
\bottomrule
\end{tabularx}

\vspace{0.25em}
\parbox{\linewidth}{\footnotesize \textit{Note.} The published SensorLLM result is reported only as background context and should not be interpreted as a fully controlled head-to-head comparison.}
\end{table}

\vspace{0.3em}

\subsection{Per-Class Analysis}

\begin{figure*}[t]
    \centering
    \includegraphics[width=\textwidth]{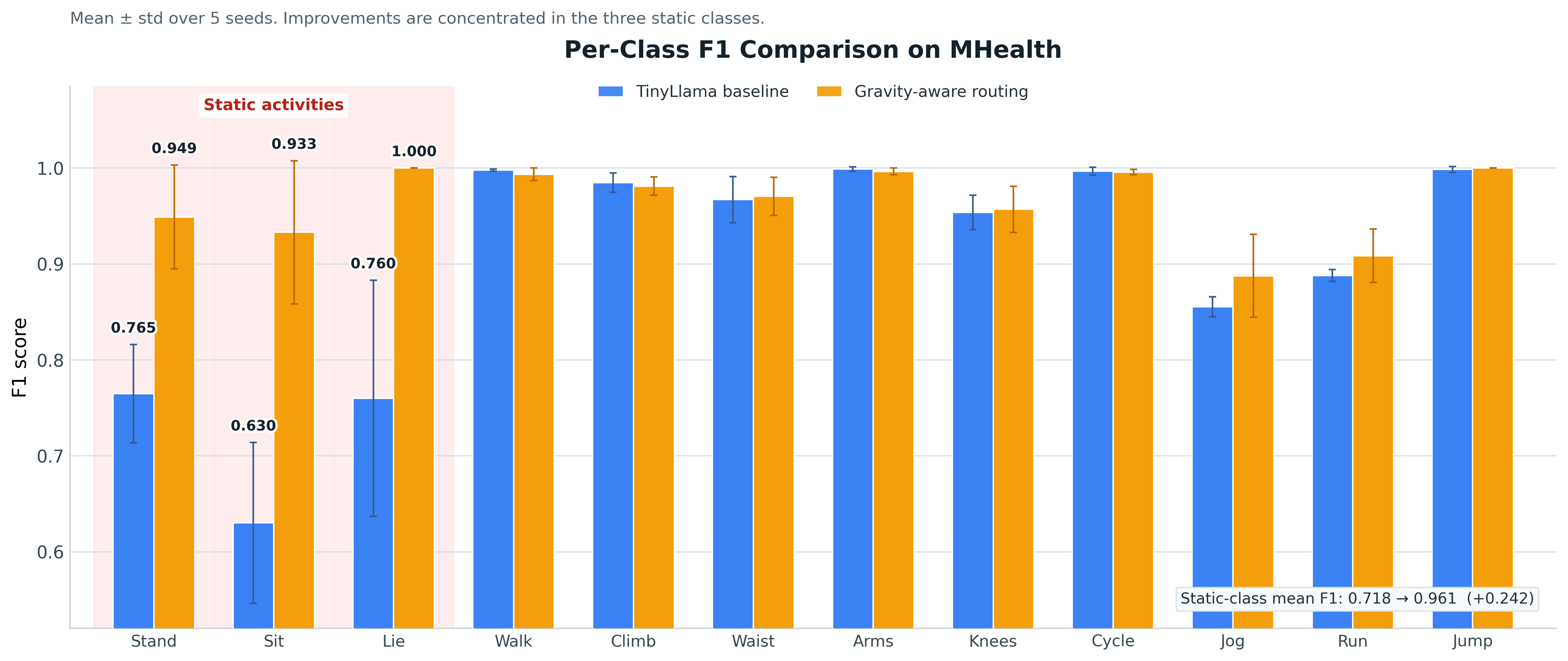}
    \caption{
    Per-class F1 comparison on MHealth. Bar heights show the mean F1 over five random seeds for the compact TinyLlama baseline and the proposed gravity-aware routing model, and error bars indicate standard deviation. The shaded region highlights the three static activity classes (standing, sitting, and lying), which are also the classes where the proposed method yields the largest gains; meanwhile, overall performance on dynamic classes remains largely stable.}
    \label{fig:fig_per_class_f1_paper}
\end{figure*}

As shown in Figure~\ref{fig:fig_per_class_f1_paper}, the performance gains are highly concentrated in the three static classes, which is exactly the target region of the proposed routing design. For standing, sitting, and lying, the compact baseline achieves F1 scores of $76.49$, $63.02$, and $76.00$, respectively, while gravity-aware routing improves them to $94.90$, $93.30$, and $100.00$. Among them, sitting shows the largest gain, reaching $+30.28$ points; standing and lying are also improved by $+18.41$ and $+24.00$ points, respectively.

This phenomenon of "gains concentrated on static classes" is important, because it shows that the proposed method does not improve all classes uniformly by generically increasing model capacity. Instead, it corrects a specific failure mode of the compact baseline, namely confusion among low-motion postural classes. In contrast, performance on dynamic classes is largely preserved, with only small fluctuations for most classes, and some non-static classes even showing modest improvements. This pattern is highly consistent with the motivation of the method and also supports the core claim of the paper: gravity-sensitive routing is especially effective for static-activity discrimination.


\subsection{Static-vs-Dynamic Grouped Results}

To summarize the above phenomenon more directly, we divide the classes into a static group (standing, sitting, lying) and a dynamic group (the other nine classes). As shown in Table~\ref{tab:static_dynamic_grouped}, after averaging over the three static classes, the compact baseline obtains a mean F1 of $71.84$, while gravity-aware routing improves it to $96.07$, corresponding to an absolute gain of $+24.23$ points. By contrast, the mean F1 over the nine dynamic classes changes only slightly from $96.01$ to $96.56$, a gain of $+0.55$ points.

This grouped result shows that the main effect of our method is not to further improve dynamic activities that are already strong, but to compensate for the clear weakness of the compact baseline on static activities. In the baseline, the static-group average is $24.17$ points lower than the dynamic-group average; after introducing routing, this gap shrinks to only $0.49$ points. In other words, the proposed head almost eliminates the "static--dynamic imbalance" that dominates the error structure of the compact baseline on MHealth.

%
\begin{table}[t]
\centering
\small
\setlength{\tabcolsep}{6pt}
\renewcommand{\arraystretch}{1.08}
\caption{Static-vs-dynamic grouped results on MHealth.}
\label{tab:static_dynamic_grouped}
\begin{tabular}{l c c c}
\toprule
Model & Static avg F1 (\%) & Dynamic avg F1 (\%) & Gap (Dynamic $-$ Static) $\downarrow$ \\
\midrule
TinyLlama baseline & 71.84 & 96.01 & 24.17 \\
Gravity routing & \textbf{96.07} & \textbf{96.56} & \textbf{0.49} \\
\bottomrule
\end{tabular}
\end{table}

\subsection{Multi-Seed Stability}

Beyond improving average performance, gravity-aware routing also substantially enhances the stability of training outcomes across random seeds. The compact TinyLlama baseline achieves $89.97 \pm 2.19$ macro-F1 and $89.28 \pm 2.39$ accuracy, while the proposed method further improves these to $96.44 \pm 0.82$ macro-F1 and $96.24 \pm 0.87$ accuracy. In other words, our method not only increases the mean of both core metrics, but also reduces their standard deviations by more than half.

\begin{figure*}[t]
    \centering
    \includegraphics[width=\textwidth]{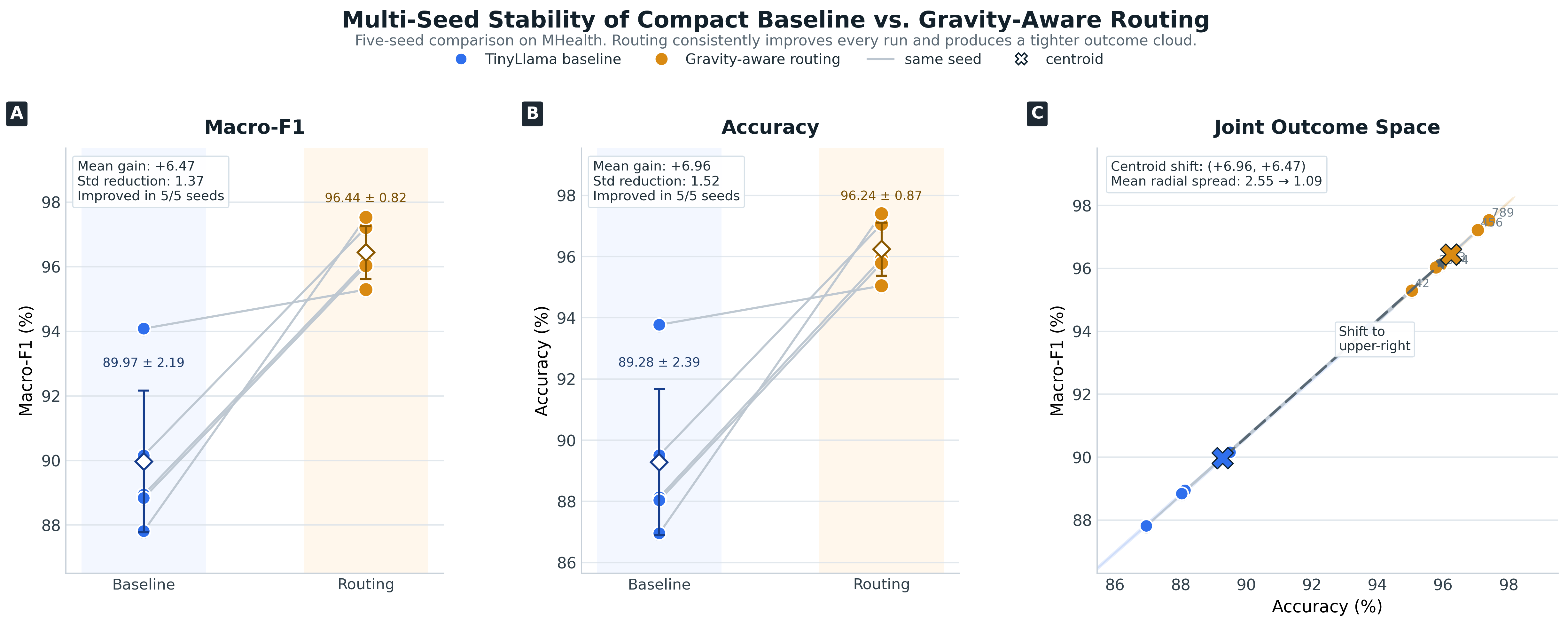}
    \caption{
    Multi-seed stability comparison between the compact TinyLlama baseline and the proposed gravity-aware routing model on MHealth. Panels A and B show paired seed trajectories for macro-F1 and accuracy, respectively, where each gray line connects the baseline and routing results obtained with the same random seed. It can be seen that gravity-aware routing consistently outperforms the compact baseline across all five random seeds. Panel C further visualizes the seed-level outcomes in the two-dimensional accuracy--macro-F1 space. Compared with the baseline, the routing model not only shifts the results as a whole toward the upper-right region, but also makes the point cloud more compact, indicating improvements in both average performance and training stability.}
    \label{fig:multiseed_stability}
\end{figure*}

Figure~\ref{fig:multiseed_stability} visualizes this phenomenon directly at the seed level. In the paired seed trajectory plots, the results for all five random seeds consistently rise from the compact baseline to gravity-aware routing, showing that the performance gain is not driven by a single favorable initialization. Specifically, across the five seeds, the macro-F1 improvement ranges from $+1.21$ to $+9.72$ points, while the accuracy improvement ranges from $+1.28$ to $+10.45$ points.

The same trend is also visible from the distributional view of seed-level outcomes. In the two-dimensional accuracy--macro-F1 space, gravity-aware routing not only shifts the results toward the upper-right region as a whole, but also makes the point cloud visibly more compact. This means that the method improves not just the "best single run," but the reproducibility of the training process itself. For compact SensorLLM-style HAR, this has strong practical significance: the proposed method does not produce strong results only occasionally, but does so more reliably across runs.

\section{Ablation Study}

To understand where the performance improvement actually comes from, we evaluate four ablation variants under the same five-seed protocol used in the main experiments. This section focuses on answering three questions: whether the gain comes merely from using more parameters, whether it comes from gravity information or only from energy information, and whether the static expert must directly receive sensor features. Unless otherwise stated, all results below are reported as mean $\pm$ standard deviation over five random seeds, as shown in Table~\ref{tab:ablation_main}.

\subsection{Is the Improvement Simply Due to More Parameters?}

The answer is no. If the performance gain mainly came from using a larger classification head, then Ablation B, which enlarges the MLP but uses neither routing nor sensor statistics, should have improved performance. However, Ablation B achieves only $84.02 \pm 0.39$ macro-F1 and $83.05 \pm 0.31$ accuracy, which is not only worse than the compact baseline at $89.97 \pm 2.19$ macro-F1 and $89.28 \pm 2.39$ accuracy, but substantially worse. This result directly rules out the simple explanation that our method performs better merely because it uses more parameters.

Ablation A provides a more informative comparison. It removes routing and directly injects sensor statistics into an MLP classification head, reaching $94.57 \pm 0.86$ macro-F1 and $94.30 \pm 0.90$ accuracy. This result clearly outperforms the compact baseline, indicating that the sensor statistics themselves do contain useful information. At the same time, however, Ablation A is still $1.87$ macro-F1 points below the full model, showing that feature injection alone is not sufficient to fully explain the final gain.

Mechanistically, these two ablations jointly suggest that the issue is not that the classification head lacks enough capacity, but that a single shared decision head cannot simultaneously accommodate the different decision requirements of static and dynamic classes. Simply increasing the size of the MLP neither explicitly separates these two decision processes nor even necessarily helps, and may instead make the decision boundary more entangled. Directly injecting sensor statistics is helpful, but without routing, the model still lacks a structured mechanism to separate posture-sensitive discrimination from general activity recognition.

\subsection{Does the Gain Come From Gravity Information, or Only From Energy Information?}

The comparison between Ablation C and the full model answers this question directly. Ablation C preserves the routing structure, but uses only the per-channel standard deviation, i.e., energy information, while removing the channel means that carry gravity- and orientation-related offset information. This variant achieves only $88.39 \pm 0.59$ macro-F1 and $87.58 \pm 0.66$ accuracy. This result is not only $8.05$ macro-F1 points below the full model, but even lower than the compact baseline itself.

This phenomenon strongly suggests that energy information alone is not sufficient to support the routing problem addressed in this paper. Low-motion static activities such as standing, sitting, and lying all tend to have weak short-term dynamics, so standard deviation alone is not enough to distinguish them reliably. The truly missing key information is the statistical cue related to gravity direction and body posture carried by the per-channel means. Once these mean features are restored, routing becomes genuinely effective.

In other words, the proposed method is not a simple energy-based router. Its core advantage is that it uses direction-sensitive statistics to determine whether a sample should rely more on the static-specialized path, thereby sending samples that truly require posture discrimination to a more appropriate decision mechanism.

\subsection{Does the Static Expert Need Direct Sensor-Feature Injection?}

Ablation D is designed precisely to answer this question. It preserves the gravity-aware router, but no longer feeds sensor features directly into the static expert. Even so, Ablation D still achieves $96.21 \pm 1.69$ macro-F1 and $96.00 \pm 1.78$ accuracy, only $0.23$ points below the full model at $96.44 \pm 0.82$ macro-F1. This shows that most of the gain of our method already comes from the routing structure itself.

Of course, the full model still performs slightly better than Ablation D, which also indicates that directly injecting sensor features into the static expert is not useless. A more reasonable interpretation is that once the router has already directed a sample to a more appropriate decision path, the additional sensor features can still provide further fine-grained correction for the static expert. That is, the router contributes the main effect, while feature injection provides a final layer of refinement for within-static discrimination.

This result is valuable because it helps clarify the relative importance of the different components in the method design. Expert decomposition and gravity-aware routing are the primary sources of improvement, while direct feature injection into the static branch is a secondary but still beneficial enhancement.

\subsection{Summary of Ablation Findings}

Taken together, these ablation results lead to a fairly consistent mechanistic conclusion. Ablation B shows that the performance gain is not driven by additional parameter count; Ablation A shows that sensor statistics are informative, but without structured routing the gain remains insufficient; Ablation C shows that energy-only routing without gravity-sensitive cues cannot solve the static-class problem; and Ablation D shows that the routing structure already explains most of the gain, while direct feature injection brings a smaller but stable additional benefit.

\begin{figure*}[t]
    \centering
    \includegraphics[width=\textwidth]{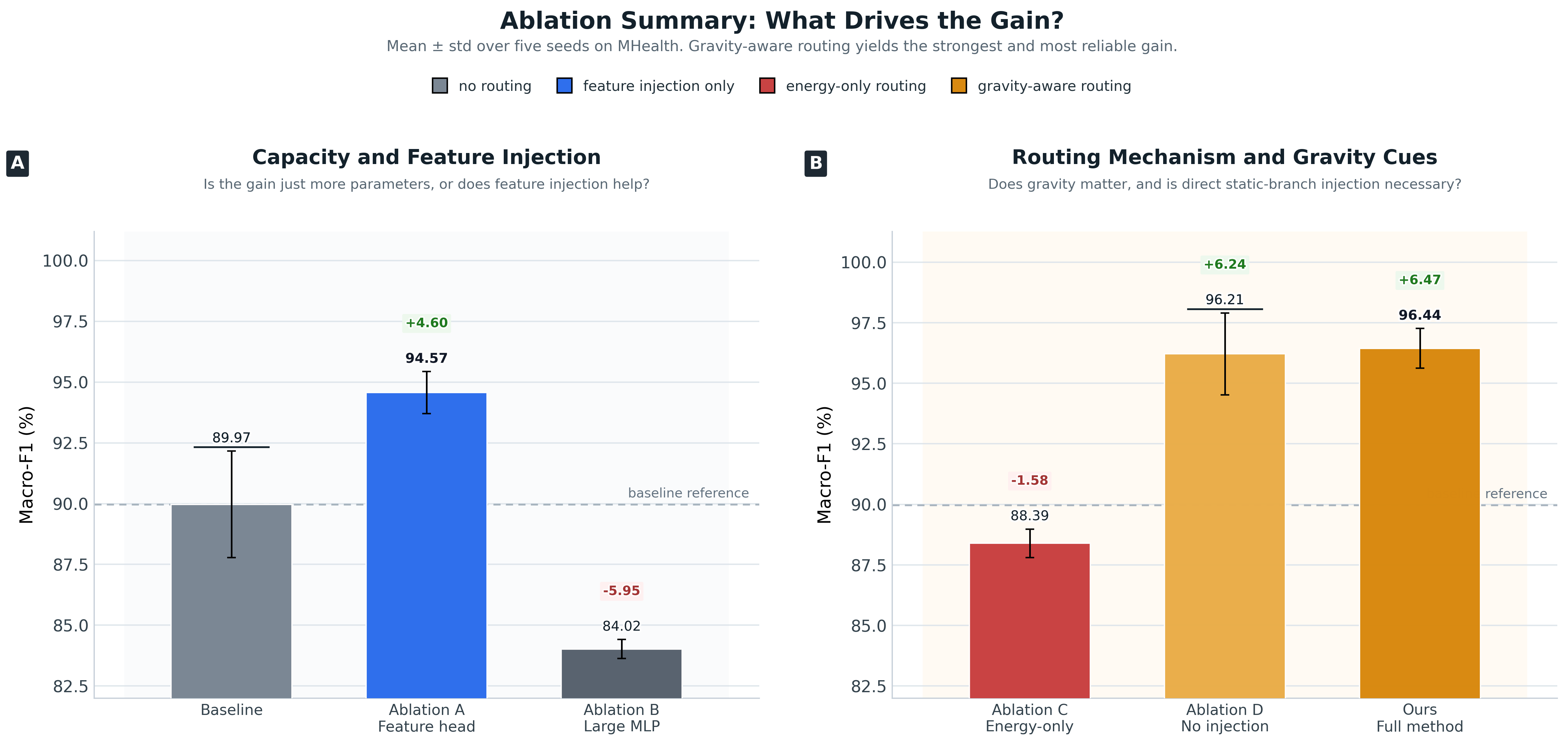}
    \caption{
    Ablation summary on MHealth. Panel A compares the compact baseline, Ablation A, and Ablation B, showing that feature injection alone is effective, whereas simply increasing the classification-head capacity does not improve performance. Panel B compares Ablation C, Ablation D, and the full model, showing that gravity-aware routing is the primary source of the performance gain, while directly injecting sensor features into the static branch provides a smaller additional benefit. Bar heights show the mean macro-F1 over five random seeds, and error bars indicate standard deviation.}
    \label{fig:ablation_summary}
\end{figure*}

The strongest evidence supporting the core claim of this paper is the contrast between Ablation C and Ablation D: both retain a routing formulation, but only the version that preserves gravity-aware statistics can approach the full model, whereas energy-only routing falls clearly behind. This directly shows that the true key to the effectiveness of the proposed method is neither extra parameters nor expertization in a generic sense, but \emph{gravity-aware hierarchical routing} itself. Figure~\ref{fig:ablation_summary} provides the corresponding visual summary.

\begin{table*}[t]
\centering
\footnotesize
\setlength{\tabcolsep}{5pt}
\renewcommand{\arraystretch}{1.10}
\caption{Ablation results on MHealth. Delta is computed relative to the compact TinyLlama baseline in macro-F1.}
\label{tab:ablation_main}
\begin{tabular}{l c c c l}
\toprule
Variant & Macro-F1 (\%) & Accuracy (\%) & Delta vs Baseline & What it tests \\
\midrule
Baseline & 89.97 $\pm$ 2.19 & 89.28 $\pm$ 2.39 & -- & compact SensorLLM reference \\
Ablation A (Feature head) & 94.57 $\pm$ 0.86 & 94.30 $\pm$ 0.90 & +4.60 & feature injection only \\
Ablation B (Large MLP) & 84.02 $\pm$ 0.39 & 83.05 $\pm$ 0.31 & -5.95 & parameter count only \\
Ablation C (Energy-only routing) & 88.39 $\pm$ 0.59 & 87.58 $\pm$ 0.66 & -1.58 & std-only routing \\
Ablation D (No feature injection) & \underline{96.21 $\pm$ 1.69} & \underline{96.00 $\pm$ 1.78} & +6.24 & routing without static injection \\
Ours & \textbf{96.44 $\pm$ 0.82} & \textbf{96.24 $\pm$ 0.87} & \textbf{+6.47} & full method \\
\bottomrule
\end{tabular}
\end{table*}

\section{Discussion}

As a supplement, we provide a compact overview of all local experiment visualizations in Appendix Figure~\ref{fig:dashboard_all}.

Our results show that the failure of compact SensorLLM-style HAR models does not occur uniformly across all classes, but is concentrated mainly on low-motion static activities such as standing, sitting, and lying. A plausible explanation is that once the backbone is compressed, the model finds it harder to preserve subtle but critical posture-sensitive differences within a shared downstream decision space. By contrast, dynamic activities usually exhibit stronger motion amplitude and more distinctive temporal patterns, and are therefore easier to keep separable even in compact models. Static activities, however, become more easily confused if the model relies mainly on short-term energy information, because their true differences are more strongly reflected in directional, postural, and gravity-related statistical structure \cite{vahaypya2018posture,lima2019haroverview}. In other words, shrinking the backbone does not merely cause an overall performance drop; it changes which sensor cues are more likely to be preserved in the final classification stage.

The effectiveness of gravity-aware routing is consistent with this explanation. The proposed method does not improve performance simply by increasing parameter count, nor does it redesign the alignment stage. Instead, it introduces a more appropriate decision structure under a compact-model setting. The router uses the per-channel mean and std to determine whether a sample should rely more on the static-specialized path, while the static expert concentrates model capacity on the subset of classes that depend most on posture discrimination. This design is effective because it aligns the model structure with the error mode of the compact baseline: dynamic activities can still be handled by a general expert, while static activities obtain finer discrimination through a specialized path that is more sensitive to gravity and orientation.

At the same time, this work should be clearly distinguished from large-scale sensor-language pretraining lines such as SensorLM and SLIP \cite{zhang2025sensorlm,chen2026slip}. Those works focus primarily on data scale, pretraining objectives, cross-task transfer, zero-shot / few-shot capability, and more general wearable representation learning \cite{zhang2025sensorlm,chen2026slip,narayanswamy2024scalingwearable,cosentino2024phllm}. In contrast, this paper studies a narrower but more specific problem: how to improve a compact Stage 2 HAR classifier under a fixed local training setup on top of an already aligned SensorLLM-style pipeline. Accordingly, this work is better understood as a lightweight post-alignment adaptation rather than a new wearable foundation model. The two directions are not simply competing alternatives, but are better viewed as complementary: large-scale pretraining can improve sensor-language representation quality, while the routing mechanism proposed here targets a downstream decision bottleneck that remains under deployment-constrained settings.

The scope of the current single-dataset first-release version also determines what we can and cannot claim. Based on the current evidence, we believe it is reasonable to say that, on MHealth, one major failure mode of compact SensorLLM-style HAR is static-activity confusion, and that gravity-aware hierarchical routing can effectively mitigate this issue under a controlled five-seed protocol. However, we cannot yet claim that the same mechanism will transfer unconditionally to other datasets, other sensor placements, other activity taxonomies, or backbones of different scales. Similarly, this paper should not be interpreted as proving that routing is universally superior to stronger representation learning or sufficient to replace large-scale pretraining. What the current version establishes is a narrower but still meaningful conclusion: in compact post-alignment HAR, when the dominant source of error is concentrated in posture-related static classes, structured specialization on the decision side can bring substantial gains.

This point is also meaningful in practice. Lightweight wearable HAR is especially important in scenarios constrained by memory, latency, power consumption, or deployment complexity, where directly using a larger backbone is often impractical \cite{lima2019haroverview,luptakova2022transformerhar}. In real systems, developers often prefer to reuse an existing aligned model and make only a small, controlled modification to the downstream classification head, rather than retrain a larger model or build an entirely new large-scale pretraining pipeline. From this perspective, the value of the proposed method lies not only in the accuracy gain itself, but also in the fact that this gain comes from a small and highly targeted backend modification while preserving most of the original pipeline structure. This makes it a practical adaptation strategy for resource-constrained wearable applications, while also leaving room for future integration with stronger sensor-language pretrained representations.

Overall, the discussion in this section supports a more restrained interpretation of the paper. We do not argue that a single routing head is sufficient to solve compact wearable HAR in general, nor do we claim that a single dataset is enough to establish broad conclusions. What we wish to emphasize instead is that the observed improvement has clear mechanistic meaning: when the main failures of compact sensor-language HAR models are concentrated on static postural classes, gravity-aware hierarchical routing provides a simple and effective corrective structure.

\section{Limitations and Future Work}

The most immediate limitation of this paper is its experimental scope. The current version validates the proposed method only on MHealth, so the evidence is better understood as a focused first report on a single dataset rather than a general conclusion about compact sensor-language HAR. Although the proposed method shows large and stable gains on this benchmark, we cannot assume that improvements of the same magnitude will automatically transfer to other datasets with different sensor placements, activity definitions, or collection protocols.

A second limitation is that the method may depend on sensor layout and channel semantics. The routing design in this paper uses per-channel mean and std extracted from the tokenizer state, and the meaning of these statistics depends to some extent on the definition of sensor axes, body placement, and modality organization. In particular, gravity-related cues are most meaningful only when channel identity can consistently preserve directional information. Therefore, for datasets with different device placements, weaker orientation consistency, or substantially different channel sets, the proposed method may require redefining the static-class subset, retuning the router, or even adopting more careful cross-dataset normalization strategies.

Third, the current focus of this paper is classification rather than broader sensor reasoning. Although more general SensorLLM-related research often takes semantic understanding and reasoning ability as an important motivation for sensor-language alignment \cite{li2025sensorllm,zhang2025sensorlm,cosentino2024phllm}, this paper does not evaluate open-ended question answering, semantic interpretation of sensor signals, cross-task transfer, or more general reasoning capabilities. Accordingly, the contribution of this work is more accurately positioned as an improvement to compact downstream HAR classification behavior, rather than a broader advance in wearable sensor reasoning.

These limitations also naturally point to future directions. First, we plan to extend the experiments to more HAR benchmarks, especially UCI, USC-HAD, and PAMAP2, to examine whether the same static-activity mechanism continues to hold under different sensor configurations and label spaces. Second, more compact LLM backbones should be tested. This paper currently uses TinyLlama as the compact Stage 2 backbone, but whether the routing design can also consistently benefit other compact LLM families is equally worth systematic investigation. More broadly, future work can explore how to combine gravity-aware routing with stronger sensor-language pretrained models, richer alignment stages, and more general reasoning-oriented evaluation protocols.

Therefore, this paper is better viewed as a starting point rather than an endpoint. It identifies a specific failure mode, provides a targeted and effective correction on one benchmark, and offers a clear direction for future systematic evaluation across multiple datasets, multiple backbones, and broader sensor-understanding tasks.

\section{Conclusion}

This paper studies a practically meaningful failure mode in compact SensorLLM-style HAR: after replacing the Stage 2 backbone with a lightweight model, recognition of low-motion static activities degrades much more severely than recognition of dynamic activities. To address this problem, we propose a gravity-aware hierarchical routing head on top of an already aligned sensor-language pipeline as a lightweight adaptation, using the per-channel mean and std to perform sample-wise routing between a static expert and a full expert. Experiments on MHealth show that this design significantly improves the compact baseline, with gains concentrated mainly on static classes such as standing, sitting, and lying, while preserving strong performance on dynamic classes. These results suggest that, under the current single-dataset setting, a small but targeted decision-side modification can effectively improve compact post-alignment HAR without changing the underlying alignment framework.

\section*{Acknowledgments}

\appendix
\section{Full Experimental Overview}

\begin{figure*}[t]
    \centering
    \includegraphics[width=\textwidth]{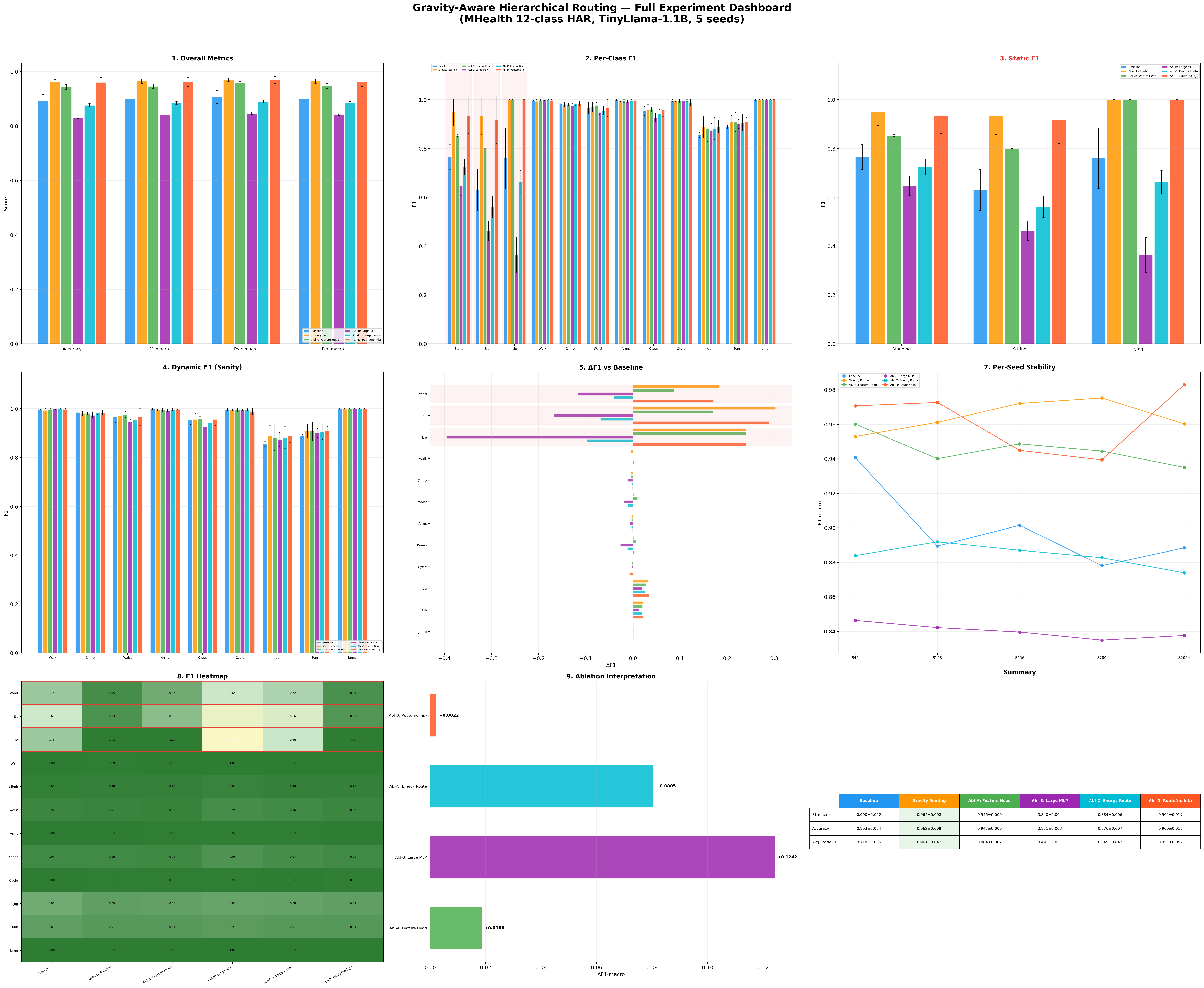}
    \caption{Comprehensive overview of the full experimental results on MHealth, summarizing overall metrics, per-class performance, static-activity performance, dynamic-class performance, multi-seed stability, and ablation results. This figure is included mainly as a compact overview in the appendix; in the main text, the key findings are instead presented as separate figures for clearer discussion.}
    \label{fig:dashboard_all}
\end{figure*}

\bibliographystyle{unsrt}
\bibliography{references}

\end{document}